\documentstyle[preprint,aps,floats,epsf,tighten,eqsecnum]{revtex}
\begin{document}
\preprint{
\parbox{1.5in}{\leftline{JLAB-THY-99-22}
                \leftline{WM-99-105}
			             \leftline{}\leftline{}\leftline{}\leftline{}}}
\title{Gravitational coupling to two-particle bound states and momentum
conservation in deep inelastic scattering}
\author{Zolt\'an Batiz$^1$ and Franz Gross$^{1, 2}$}
\address{$^1$Department of Physics, College of William and Mary,
 Williamsburg, VA  23185 \\ $^2$Thomas Jefferson National Accelerator Facility, 
Newport News, VA 23606}
\date{\today}
\maketitle

\begin{abstract}
The momentum conservation sum rule for deep inelastic scattering (DIS) from
composite  particles is investigated  using the general theory of relativity. 
For two $1+1$ dimensional examples, it shown that covariant theories
automatically satisy the DIS momentum conservation sum rule provided the bound
state is covariantilly normalized.  Therefore, in these cases the two DIS sum
rules for baryon conservation and momentum conservation are equivalent.  
\end{abstract}

\def\s{\ \! / \! \! \! \!}

\widetext
\section{Introduction}

In the parton model of deep inelastic scattering (DIS) from a bound state with
$N$ valence quarks, one encounters two sum rules:
\begin{eqnarray}
\sum_i n_i \int_0^1 f_i(x) dx  &&=\sum_i n_i\label{eq1aa}\\
\sum_i\int_0^1 x f_i(x) dx  &&= 1\, , \label{eq1}
\end{eqnarray}
where $f_i(x)$ is the probability that the quarks of flavor $i$ will have
momentum fraction $x$.  The first sum rule corresponds to baryon (or charge)
conservation and the second to the conservation of the momentum fraction.  In a
previous work\cite{ZFBG1}, we showed, in the context of a covariant model for
the bound state, that the first sum rule is a consequence of the normalization
condition for the covariant bound state wave function, and hence is
automatically satisfied in any covariant model.  

In parton models, the two relations (\ref{eq1}) are usually considered to be
independent constraints.  This could pose a problem for covariant descriptions
of DIS because in this formalism the normalization condition is the only
constraint which can be imposed on the covariant wave function.  It is
therefore unclear where the additional momentum conservation constraint could
come from.  This suggests that either covariant models of DIS are inconsistent,
or that the two constraints (\ref{eq1}) are somehow not independent and in fact
are {\it both\/} a consequence of the normalization condition.  This puzzle
provided the initial motivation for this work.  We will return to discussion of
this issue in the concluding section.

In the main part of this paper we turn to another question related to the above
issue:  ``How can we prove, in the context of a covariant treatment of
two body bound states, that a static gravitational field couples only to the
total mass of the bound state?''  In the context of a field theory this
requires proving that the coupling of a long wave length graviton to a
composite system has the same structure as its coupling to an elementary
particle, with the mass replaced by the total mass of the bound state.  In
this paper we will show, for two simple 1+1 dimensional models, that the bound
state normalization condition is sufficient to insure that is indeed the
case.  

These conservation laws are a consequence of the symmetries of the external
field with which the system interacts.  For example, the deuteron, which is
the bound state of a proton and a neutron, participates in the electromagnetic
interaction with a $U(1)$ symmetry.  The electromagnetic field couples to
each of the constituents separately but if the radiation has a large
wavelength, it ``sees'' only the total charge. Consequently one should be able
to compute the total charge of the composite system from the density functions
predicted from the bound state model, so one gets  a constraint on the bound
state. As is well known, this constraint is identical to the bound state
normalization condition.  

To see how this comes about in the context of the gravitational interaction
(related to momentum conservation), we go back to the first principles and
consider the interaction of a composite particle in a gravitational field.
The requirement that the gravitational field couple to the total bound state
mass leads to a constraint which at first seems to be different from the
the wave function normalization condition, but we prove that they are
compatible.
 
We start in the next section by reviewing the properties of the gravitational
interaction with scalar particles and develop the general framework of our
approach. Then we apply our method to the study of some two-particle bound
states in $1+1$ dimensions. In Sec.~III we study bound states of two scalars,
and in Sec.~IV bound states of a spin $1/2$ fermion and a scalar.  Some
conclusions are given in Sec.~V.

\section{Gravitational Interactions with scalar particles}

In this section we review the most basic properties of 
the lowest order graviton-scalar interaction. After we obtain the Feymnan rules
of  two cubically interacting scalar 
fields in the environment of a gravitational field, we 
investigate some Feynman diagrams with a single 
external graviton line to illustrate momentum 
conservation in some specific cases. 
We make use of the gravitational Ward-Takahashi identity, 
which we first prove by using our 
previously obtained Feynman rules and prove using a more general field
theory  argument.

This section is meant to give the general framework of this paper, 
including the machinery of the bound
state-graviton interaction, that will be used 
in the next sections to derive the constraint imposed on the bound state wave 
functions by momentum conservation.

\subsection{Gravitational Interaction of Scalar Particles}

Consider two interacting 
scalar fields $\Phi$ and $\phi$.  We will assume $\Phi$ is 
charged and that $\phi$ is a neutral, self conjugate field. 
A gravitational ``interaction'' is added 
if the flat metric tensor $\eta^{\mu \nu}=diag[1,-1,-1,-1]$ is replaced by
an  arbitrary metric $g^{\mu \nu}(x)$,
\begin{equation}
g^{\mu \nu}(x)=\eta^{\mu \nu} + 2h^{\mu \nu}(x)\, , \label{eq2}
\end{equation}
where $h^{\mu \nu}(x)$ is the gravitational field. (Note that our metric
differs in sign and that $h^{\mu \nu}$ differs by a factor of 2
from that used by Weinberg\cite{WEINBERG}). The four-volume element
$d^4 x$ is also replaced by  
\begin{equation}
\int d^4x \to \int d^4 x \sqrt{-g}\, , \label{eq3}
\end{equation}
where $g$ is the determinant of the 
{\it covariant\/} metric tensor $g_{\mu \nu}$. [Recall that, in curved space,
the covariant and contravariant metric tensors are not the same.] 

For simplicity, the $\Phi \Phi \phi$ interaction 
is assumed to be a Yukawa-type, so the 
Lagrangian reads:
\begin{equation}
{\cal L}= g^{\mu \nu} 
\partial_{\mu} \Phi^{\dagger} 
\partial_{\nu} \Phi -M^2 \Phi^{\dagger} \Phi
+\frac{1}{2} g^{\mu \nu} \partial_{\mu} \phi 
\partial_{\nu} \phi- \frac{1}{2} m^2 \phi^2
- \lambda \phi \Phi^{\dagger} \Phi,
\label{1eq1}
\end{equation}
The kinetic part of the Lagrangian of the scalar fields naturally
involves  the covariant components of the 
derivatives of the fields, rather than 
the contravariant ones.   Since the fields are scalars, their plain derivatives
are therefore identical to their covariant derivatives, and hence for scalar
fields the  affine connections
\begin{equation}
\Gamma^{\alpha}_{\mu \nu}=\frac{1}{2}g^{\alpha \beta}(g_{\beta \mu , \nu}+
g_{\nu \beta, \mu}-g_{\mu \nu,\beta})
\label{1eq1a}
\end{equation}
(where $g_{\mu \nu,\beta}=\partial g_{\mu \nu}/\partial x^\beta$) usually needed
to convert normal derivatives in covariant derivatives, will play no role.  This
greatly simplifies the discussion of the gravitational interactions of scalar
fields.  Later, when we discuss spin 1/2 fields in Sec.~IV, we will have to
deal with the greater complexity introduced by the affine connections.

In four dimensions, the total action for this scalar theory is taken to
be\cite{WEINBERG,KAKU,OHANIAN1}  
\begin{eqnarray}
S=\int d^4 x \sqrt{-g} \left[{\cal L} 
+\frac{1}{2\kappa} R \right]=S_f+S_g\label{1eq4}
\end{eqnarray}
where  
$$\kappa=\frac{8 \pi G}{c^4}$$
and $R$ is the curvature scalar of the space-time 
continuum equal to  $g^{\mu \nu} R_{\mu \nu}$, where $R_{\mu \nu}$ is the
Ricci tensor
\begin{equation}
R_{\mu \nu}=-\Gamma^{\alpha}_{\mu \nu , \alpha} + 
\Gamma^{\alpha}_{\mu \alpha, \nu}+ 
\Gamma^{\lambda}_{\mu \alpha} 
\Gamma^{\alpha}_{\lambda \nu} -
\Gamma^{\lambda}_{\mu \nu} 
\Gamma^{\alpha}_{\lambda \alpha}\, .
\label{0eq3}
\end{equation}
In the special case of the 
flat space, $R=0$, and the usual 
Euler-Lagrange equations result 
from the application of Hamilton's principle to 
the fields $\phi$ and $\Phi$:
\begin{eqnarray}
\eta^{\mu \nu} \partial_{\mu} 
\partial_{\nu} \phi +m^2 \phi +\lambda 
\Phi^{\dagger} \Phi&&=0 \nonumber \\
\eta^{\mu \nu} \partial_{\mu} 
\partial_{\nu} \Phi +M^2 \Phi +\lambda
\Phi \phi &&=0\, .
\label{1eq5}
\end{eqnarray}   
Quantizing the fields using the path integral formalism 
gives the  Feynman rules:
\begin{eqnarray}
-i{\cal V}_3=&&-i\lambda \nonumber\\
-i\Delta_M(p)=&&\frac{-i}{M^2-p^2-i \epsilon}\, ,  
\label{1eq6}
\end{eqnarray}
where ${\cal V}_3$ is the Feynman rule of the $\Phi \Phi \phi$ vertex (Fig.
1), and $\Delta_M$ is the propagator for the $\Phi$ field. [A
similar factor, $\Delta_m$, is the propagator  for the neutral field $\phi$.] 
In our Feynman diagrams, the $\Phi$ field is represented by the thick 
solid line and the $\phi$ field is 
represented by the thin solid line.  We obtain the usual equations of 
motion and the usual Feynman rules for 
the limiting case of a flat space.

\begin{figure}
\centerline{\epsfxsize=0.8in\epsffile{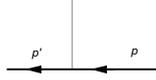}}
\caption{$\Phi \Phi \phi$ interaction.}\label{one}
\end{figure}

In four dimnensions the action (\ref{1eq4}) also yields Einstein's
equations for the field
$h^{\mu \nu}$ defined in Eq.~(\ref{eq2}).  Hamilton's principle for the 
field $h^{\mu \nu}$ is
\begin{equation}
\frac{\delta}{\delta h^{\mu \nu}} 
\left[\sqrt{-g}\left({\cal L} +\frac{1}{2\kappa} R\right)\right]=0\, ,
\label{1eq7}
\end{equation}
and yields the following Euler-Lagrange equations for the 
gravitational field
\begin{equation}
\frac{1}{\kappa}\left\{R_{\mu \nu}-\frac{1}{2}\,g_{\mu \nu} R\right\} +
{\cal T}_{\mu\nu}=0\, .
\label{1eq9}
\end{equation}
Here the total energy-momentum tensor ${\cal T}_{\mu \nu}$ is the sum of
contributions from the fields $\phi$ and $\Phi$, and an interaction part:
\begin{eqnarray}
&&{\cal T}_{\mu \nu}=T_{\mu \nu}
(\phi)+T_{\mu \nu}(\Phi)+
T_{\mu \nu}(\phi, \Phi) \nonumber \\
&&T_{\mu \nu}(\phi)=\frac{1}{2}
(\partial_{\mu} \phi \partial_{\nu} \phi + 
\partial_{\nu} \phi \partial_{\mu} \phi)
 -\frac{1}{2}g_{\mu \nu}
(g^{\alpha \beta} 
\partial_{\alpha} \phi \partial_{\beta} \phi-
m^2 \phi^2) 
\nonumber \\
&&T_{\mu \nu}(\Phi)=\partial_{\mu} 
\Phi^{\dagger} \partial_{\nu} 
\Phi + \partial_{\nu} 
\Phi^{\dagger} \partial_{\mu} \Phi
-g_{\mu \nu}(g^{\alpha \beta}
\partial_{\alpha} \Phi^{\dagger} 
\partial_{\beta} \Phi-M^2 
\Phi^{\dagger} \Phi) \nonumber \\
&&T_{\mu \nu}(\phi, \Phi)
=\lambda g_{\mu \nu} \Phi^{\dagger} 
\Phi \phi\, .
\label{1eq8}
\end{eqnarray}  

The ${\cal T}_{\mu \nu}$ term in Eq.~(\ref{1eq9}) can be derived easily once the
variation of $\sqrt{-g}$ in terms of $\delta g^{\mu\nu}=2h^{\mu\nu}$
has been calculated.  We will carry out this derivation in an arbitrary
number of (integral) dimensions $d$.  First, exploit the identity
\begin{equation}
g^{\mu \nu} g_{\mu \nu} =d 
\label{1eq10c}
\end{equation}
to obtain the relation 
\begin{equation}
g^{\mu \nu}h_{\mu \nu}=-g_{\mu \nu}h^{\mu \nu}\, .
\end{equation}
This can also be conveniently rewritten
\begin{equation}
h_{\alpha \beta}=-\frac{1}{d}\,g_{\alpha \beta}\,g_{\mu \nu}h^{\mu \nu}\, .
\end{equation}
From this last relation, and the definition of $g$ in $d$ dimensions 
\begin{equation}
g=N\epsilon_d^{\alpha\beta\cdots}\epsilon_d^{\alpha'\beta'\cdots}
\,g_{\alpha \alpha'}g_{\beta \beta'}\cdots\, ,
\end{equation}
where $\epsilon_d$ is the antisymmetric symbol in $d$ dimensions and $N$
is a sutiable normalization constant, it follows
that
\begin{eqnarray}
\delta g &&=
2Nd\,\epsilon^{\alpha\beta\cdots}\epsilon^{\alpha'\beta'\cdots}
\,h_{\alpha \alpha'}g_{\beta \beta'}\cdots
\nonumber\\
&&=- 2g\,g_{\mu \nu}h^{\mu \nu}\, ,
\end{eqnarray}
and therefore
\begin{equation}
\delta \sqrt{-g}=-\sqrt{-g}\,
g_{\mu \nu} h^{\mu \nu}. 
\label{1eq10d}
\end{equation}
On the other hand
\begin{equation}
\frac {\delta {\cal L}} 
{\delta h^{\mu \nu}}=
2\partial_{\mu} 
\Phi^\dagger \partial_{\nu} \Phi+ \partial_{\mu} 
\phi \partial_{\nu} \phi. 
\label{1eq10f}
\end{equation}
Putting Eqs. (\ref{1eq10d}) and (\ref{1eq10f}) together, and symmetrizing 
(\ref{1eq10f}), we see that:
\begin{equation}
\delta \int d^4 x \sqrt{-g}\, {\cal L}= 
\int d^4 x \sqrt{-g} \,{\cal T}_{\mu \nu}
\,h^{\mu \nu}\, . 
\label{1eq10g}
\end{equation}

The first two terms in Eq.~(\ref{1eq9}) arise from the variation of $S_g$ in
four dimensions\cite{WEINBERG,OHANIAN1}
\begin{equation}
\delta \left(\frac{1}{2 \kappa} \int d^4 x \sqrt{-g} R \right) =
\frac{1}{\kappa} \int d^4 x \sqrt{-g}\,(R_{\mu \nu}-
\frac{1}{2} g_{\mu \nu} R) 
\,h^{\mu \nu}\, . 
\label{1eq10e}
\end{equation} 
Setting the combined variations (\ref{1eq10g}) and (\ref{1eq10e}) to zero
gives the field equations (\ref{1eq9}).  These are Einstein's equations 
for the gravitational ``interaction''\cite{WEINBERG} and hence the action 
(\ref{1eq4}) gives the correct classical relation for the gravitational field
in  addition to the right flat space limit.

\begin{figure}
\centerline{\epsfxsize=2in\epsffile{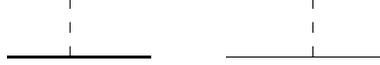}}
\caption{$\Phi \Phi$-graviton and $\phi
\phi$-graviton interaction.}\label{two}
\end{figure}

It will turn out that Eq.~(\ref{1eq10e}) must be modified for the
applications to 1+1 dimensional space studied in this paper (see the
discussion in Subsection B below).  This leads to a modification of the field
equation (\ref{1eq9}) in 1+1 dimensions.  However, the derivation of the
coupling of the gravitational field to matter, given in Eq.~(\ref{1eq10g}),
is, as we have seen, independent of dimension, and we can derive the Feynman
rules which describe the coupling of a graviton to matter directly from it. We
will only consider the gravitational interaction to lowest order (weak field
limit) and will represent the gravitational  interaction by a dashed line.  The
Feynman rules, besides the ones given in Eqs.~(\ref{1eq6}),  contain two matter
field-graviton vertices (Fig \ref{two}). 
The Feynman rule of these vertices is:
\begin{eqnarray}
-i {\cal J}_M^{\mu \nu}(p',p)=i
\left[p^{\mu}
p^{\prime \nu}+p^{\nu}p^{\prime \mu}
+\eta^{\mu \nu}(M^2-p \cdot p^{\prime})\right]\, . 
\label{1eq10}
\end{eqnarray}
The $\phi \phi$-graviton vertex has the same form, as 
expected from the fact that charged and uncharged particles interact in the
same way with a gravitational field. 

This Feynman 
rule can be derived from the flat space limit of Eq. (\ref{1eq10g})
\begin{equation}
\delta \left(i \int d^4 x \sqrt{-g} {\cal L} \right)
\simeq i \int d^4 x  {\cal T}_{\mu \nu}(x)
h^{\mu \nu}(x) \, .
\label{1eq10i}
\end{equation} 
Sandwiching this result between 
the incoming and outgoing $\Phi$ (or $\phi$) 
states gives
\begin{equation}
\delta \left(i<p^{\prime}|\int d^4 x 
\sqrt{-g} {\cal L} |p> \right)
\simeq i\int d^4 x <p^{\prime}| {\cal T}_{\mu \nu}(x) |p>
h^{\mu \nu}(x)\, . 
\label{1eq10j}
\end{equation} 
Reducing this expression gives
\begin{eqnarray}
&&\delta \left(i<p^{\prime}|\int d^4 x 
\sqrt{-g} {\cal L} |p> \right) \nonumber\\
&&\qquad\qquad\simeq  
i\left[p_{\mu}p'_{\nu}+ 
p_{\nu}p'_{\nu} + \eta_{\mu \nu}
(M^2-p \cdot p^{\prime})\right] 
\int d^4 x h^{\mu \nu}(x) 
e^{i(p'-p) \cdot x} \nonumber \\ 
&&\qquad\qquad=i \left[p_{\mu}p'_{\nu}+
p_{\nu}p'_{\nu}+
\eta_{\mu \nu}(M^2-p \cdot p^{\prime})\right]
{\tilde h}^{\mu \nu}(p^{\prime}-p)\, ,  
\label{1eq10k}
\end{eqnarray} 
where ${\tilde h}^{\mu \nu}(p^{\prime}-p)$ is 
the Fourier transform of the gravitational field.  Hence the coupling
given in Eq. (\ref{1eq10}) [which is the coefficient of the 
${\tilde h}^{\mu\nu}$ term in $\delta{\cal L}$] emerges.   Although this 
derivation assumes that the gravitational field  is not quantized, Eq.
(\ref{1eq10}) is correct  even in the case of quantized gravitational fields,
as one will  see in Subsection C.  

\begin{figure}
\centerline{\epsfxsize=1in\epsffile{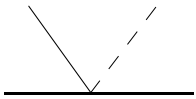}}
\caption{$\Phi \Phi \phi$-
graviton vertex}\label{three}
\end{figure}

The model likewise involves a $\Phi \Phi 
\phi$-graviton coupling (Fig.\ref{three}),
its Feynman rule being:
\begin{equation}
-i{\cal V}_{\mu \nu}^3= i \lambda
\eta_{\mu \nu}.
\label{1eq11}
\end{equation}
The vertex ${\cal V}_{\mu \nu}$ has 
a physical interpretation:
it is the "weight" 
of the vertex ${\cal V}$. In 
the non-quantized theory of the 
point-like particles, 
the gravity couples to 
the energy-momentum tensor, 
which may have 
a part coming from an interaction. 
Our interpretation is 
consistent with the limiting case. 
Note the obvious, model independent generalization:
\begin{equation}
-i {\cal V}_{\mu \nu}^3=
i\eta_{\mu \nu}{\cal V}^3,
\label{1eq11a}
\end{equation} 
All these vertices involving 
gravitons are symmetric in the Lorentz 
indices, and so is the external 
gravitational field $h^{\mu \nu}$.   

In addition to these interactions, the 
theory involves a graviton propagator, 
graviton-graviton interactions
and several-graviton -matter 
field vertices, which we do not display. 
Because we are 
interested in first order gravitational 
interactions between a bound state of two 
$\Phi$ particles and an external 
(non-quantized) weak 
gravitational field, 
we can safely ignore these undisplayed rules. 

In the next subsection we extend our discussion of the field equations 
to 1+1 dimensional space and describe the constraints on the gravitational
couplings which emerge from gauge invariance.

\subsection{Gravitational Coupling and Gauge Invariance}

Because the gravitational action does not contain a mass term, scattering
amplitudes at any order are expected to be  gauge invariant.  [The cosmological
term\cite{OHANIAN1} would destroy  this symmetry, but if it is non-zero it is
at least very small, so gauge invariance is either exact or a very good
approximation.]  In four dimensions this follows from Eq.~(\ref{1eq9}); the
covariant divergence of the $1/\kappa$ term vanishes identically, insuring that
the energy momentum tensor is conserved.  In addition, the gravitational
trace of Eq.~(\ref{1eq9}) yields the result
\begin{equation}
\left(\frac{d}{2}-1\right)\,R = \kappa \, g^{\mu\nu}{\cal T}_{\mu\nu}\, .
\label{1eq11aa}
\end{equation} 
Note that this equation cannot be correct in $d=2$ dimensions, where the
LHS would be zero and the RHS non-zero.  This is one indication that the
Eq.~(\ref{1eq9}) is not correct in two (i.e. 1+1) dimensions.    

For our purposes it is sufficient to find the correct equation in the weak
gravitational field limit.  Eq.~(\ref{1eq9}) reduces to
\cite{OHANIAN1}
\begin{eqnarray}
&&\partial^2 h^{\mu \nu} -
\partial_{\lambda} \partial^{\nu} h^{\mu \lambda}-
\partial_{\lambda} \partial^{\mu} 
h^{\nu \lambda}  +\eta^{\mu \nu} 
\partial_{\lambda} \partial_{\sigma} 
h^{\lambda \sigma} + \partial^{\mu} \partial^{\nu} h
-\eta^{\mu \nu} \partial^2 h=-\kappa {\cal T}^{\mu
\nu}\, , \label{linear1}
\end{eqnarray} 
where $\partial^2=\partial_{\lambda} 
\partial^{\lambda}$ and $h=\eta_{\alpha \beta} h^{\alpha \beta}$.  As
expected, the trace of the LHS is zero in 2 dimensions (to lowest order it is
sufficient to calculate this trace by contracting with $\eta_{\mu\nu}$). 
However, as discussed by Ohanian and Ruffini\cite{OHANIAN1}, the following
replacement 
\begin{eqnarray}
h^{\mu\nu}= \bar{h}^{\mu\nu} -\frac{1}{2}(1-a) \;\eta^{\mu\nu}\bar{h}\, ,
\end{eqnarray} 
transforms Eq.~(\ref{linear1}) to
\begin{eqnarray}
&&\partial^2 \bar{h}^{\mu \nu} -
\partial_{\lambda} \partial^{\nu} \bar{h}^{\mu \lambda}-
\partial_{\lambda} \partial^{\mu} 
\bar{h}^{\nu \lambda}  +\eta^{\mu \nu} 
\partial_{\lambda} \partial_{\sigma} 
\bar{h}^{\lambda \sigma} + a\left(\partial^{\mu} \partial^{\nu} \bar{h}
-\eta^{\mu \nu} \partial^2 \bar{h}\right) =-\kappa {\cal T}^{\mu
\nu}\, , \label{linear2}
\end{eqnarray} 
showing that the coefficient $a$ is arbitrary (except it cannot be 1/2 because
this choice would force $\eta_{\mu\nu}h^{\mu\nu}=0$).  Since
$h^{\mu\nu}$ is symmetric, the divergence of the LHS of
Eq.~(\ref{linear2})  with respect to either index is zero, and the 
stress energy tensor is conserved for any value of $a$
\begin{equation}
\partial_\mu {\cal T}^{\mu \nu} = 0 = \partial_\mu {\cal T}^{\nu \mu}\, .
\label{gauge1}
\end{equation}
This is analogous to the conservation of electromagnetic current.  However,
the trace condition obtained from Eq.~(\ref{linear2})  now depends on
$a$ and is not zero unless $d=2$ and $a=1$
\begin{eqnarray}
&&(1+a-ad)\partial^2 \bar{h} +(d-2)
\partial_{\mu} \partial_{\nu} \bar{h}^{\mu \nu}  =
-\kappa \,\eta_{\mu\nu} 
{\cal T}^{\mu \nu} =
(1-a) \partial^2 \bar{h}\, , \label{linear3}
\end{eqnarray} 
where the last form of the equation holds in $d=2$ dimensions.  We will not
persue the formal development of the 2 dimensional theory further.  We will
imagine that the gravitational field in 2 dimensions is given by
Eqs.~(\ref{linear2}) and (\ref{linear3}) with $a\ne 1$.

Because of the symmetry under interchange of indices, the gravitational gauge
invariance is described by a $U(1)\times U(1)$ group and we may check
the  gauge invariance for the case of one 
index only. In momentum space the requirements (\ref{gauge1}) therefore
become, in the linear approximation,
\begin{equation}
q_\mu {\cal T}^{\mu \nu} = 0 \, ,
\label{gauge2}
\end{equation}
where $q$ is the momentum of the graviton. 
Let us proceed by verifying gravitational gauge invariance 
for some simple cases.

First consider the graviton-$\Phi \Phi$ vertex given in
Eq.~(\ref{1eq10}) and shown in  Fig.~2. If neither of the $\Phi$ mesons is
on-shell, the divergence of the  interaction term (\ref{1eq10}) gives the
following Ward-Takahashi identity:
\begin{equation}
q_{\mu}{\cal J}^{\mu \nu}_M(p',p)= 
p^{\nu} \Delta_M^{-1}(p^{\prime})-p^{\prime \nu} \Delta_M^{-1}(p)\, ,
\label{1eq12}
\end{equation}
where $p'$ and $p$ are the final and initial momentum of the particles and
$q=p'-p$.  If the particles are on-shell, the 
inverse propagators vanish, so 
the vertex is gauge invariant. 
The graviton-$\phi\phi$ vertex is 
similarly gauge invariant for 
on-shell $\phi$ particles, and 
its Ward-Takahashi identity 
is the same, with $M$ replaced by $m$.
A more complete description 
of the graviton-scalar 
interaction can be found in \cite{FEYNMAN}.

The gravitational Ward-Takahashi 
identities are important 
relations needed to verify the gauge 
invariance of the higher order diagrams.
Therefore, before going further in 
checking some other diagrams,
we will devote the next subsection to a model- 
independent proof of this important identity.

\subsection{Ward-Takahashi Identities}
In this subsection we derive 
the gravitational Ward-Takahashi 
identity for any scalar field 
in a model independent way.
A derivation has been given by West \cite{WEST}, but here we follow a different
path based on ideas borrow from references \cite{SAVKLI} and \cite{PESKIN}, 
where the Ward-Takahashi identity has been derived for QED and QCD.

We start with the field theoretic matrix element:
\begin{eqnarray}
(2 \pi)^d  &&\delta^{(d)}(p+q-p^{\prime}) 
\Delta_M (p^{\prime}){\cal J}_{\mu \nu}
(p^{\prime}, p) \Delta_M(p) \nonumber\\ 
&&=\frac{1}{2} \int d^d x e^{-ip^{\prime} \cdot x} 
\int d^d y e^{iq \cdot y} 
\int d^d z e^{ip \cdot z} 
<0| T \left(\Phi (x) { T}_{\mu \nu}(y) 
\Phi^{\dagger}(z) \right)|0>\, , 
\label{Beq1}
\end{eqnarray}
where $T_{\mu \nu}(y)=T_{\mu \nu}(\Phi(y))$ was given in Eq.~(\ref{1eq8}), and
the factor of one-half on the RHS insures that the ${\cal J}_{\mu \nu}$ given on
the LHS agrees with our definition (\ref{1eq10}). Translating the fields (where
${\hat p}$ is the momentum operator)
\begin{eqnarray}
\Phi^{\dagger}(z)=&&e^{-i {\hat p} \cdot \eta} 
\Phi^{\dagger}(z-\eta) e^{i {\hat p} \cdot \eta}
\nonumber \\
{T}_{\mu \nu}(y) =&& e^{-i {\hat p} \cdot \eta} 
{T}_{\mu \nu}(y-\eta) e^{i {\hat p} \cdot \eta}\nonumber \\
\Phi(x) =&& e^{-i{\hat p} \cdot \eta} \Phi(x-\eta) e^{i {\hat p} \cdot \eta}
\label{Beq2}
\end{eqnarray}
changes the time ordered product to:
\begin{equation}
T\left(\Phi(x)  
{T}_{\mu \nu}(y) \Phi^{\dagger}(z) \right)=
T\left( e^{-i {\hat p} \cdot \eta} \Phi(x-\eta)  
{ T}_{\mu \nu}(y-\eta) \Phi^{\dagger}(z-\eta) 
e^{i {\hat p} \cdot \eta} \right)\, .
\label{Beq3}
\end{equation}
Letting $\eta=(x+z)/2$, changing variables
\begin{eqnarray}
x-\eta&&=\frac{1}{2}(x-z) \rightarrow \frac{1}{2}\zeta \nonumber \\
y-\eta &&\rightarrow y \, ,
\label{Beq4}
\end{eqnarray}
and integrating over $\eta$, transforms (\ref{Beq1})  into 
\begin{eqnarray}
\Delta_M(p^{\prime}) &&
{\cal J}_{\mu \nu}(p^{\prime}, p)\Delta_M(p) \nonumber\\
=&&\frac{1}{2}\int d^d \zeta e^{-i(p^{\prime} +p)\cdot \frac{1}{2}\zeta} 
\int d^d y e^{i q \cdot y}
<0| T \left( \Phi\left({\textstyle\frac{1}{2}}\zeta\right) 
{T}_{\mu \nu}(y) \Phi^{\dagger}\left(-{\textstyle\frac{1}{2}}\zeta\right) 
\right)|0>\, ,
\label{Beq5}
\end{eqnarray}
where we have used the translational invariance of the vacuum.
Dotting this equation with $q$, and integrating by parts, gives
\begin{eqnarray}
q^{\mu} \Delta_M(p^{\prime})&& {\cal J}_{\mu \nu}(p^{\prime}, p)\Delta_M(p)
\nonumber\\ =&& \frac{i}{2}\int d^d \zeta 
e^{-i(p^{\prime}+p) \cdot \frac{1}{2}\zeta} \int d^d y e^{i q \cdot y}
\frac{\partial}{\partial y_{\mu}} <0| T\left(
\Phi\left({\textstyle\frac{1}{2}}\zeta\right) {T}_{\mu\nu}(y) 
\Phi^{\dagger}\left(-{\textstyle\frac{1}{2}}\zeta\right) \right)|0>\, .
\label{Beq7}
\end{eqnarray}
Since ${T}_{\mu \nu}$ is divergenceless, the divergence of the temporal product
in (\ref{Beq7}) reduces to
\begin{eqnarray}
&&\frac{\partial}{\partial y_{\mu}} T \left( \Phi
{T}_{\mu \nu} \Phi^{\dagger} \right)\nonumber\\
&&=\delta(y_0-{\textstyle\frac{1}{2}}\zeta_0) \left( \theta 
(\zeta_0) \left[ {T}_{0 \nu}(y), 
\Phi\left({\textstyle\frac{1}{2}}\zeta\right) \right] 
\Phi^{\dagger}\left(-{\textstyle\frac{1}{2}}\zeta\right)+
\theta(-\zeta_0) \Phi^{\dagger}\left(-{\textstyle\frac{1}{2}}\zeta\right)
\left[ { T}_{0 \nu}(y), \Phi\left({\textstyle\frac{1}{2}}\zeta\right) 
\right] \right) \nonumber \\
&&+\delta(y_0+{\textstyle\frac{1}{2}}\zeta_0)
\left( \theta(\zeta_0) \Phi\left({\textstyle\frac{1}{2}}\zeta\right) \left[
{ T}_{0 \nu}(y), \Phi^{\dagger}\left(-{\textstyle\frac{1}{2}}\zeta\right)
\right]+\theta(-\zeta_0) \left[{ T}_{0 \nu}(y), 
\Phi^{\dagger}\left(-{\textstyle\frac{1}{2}}\zeta\right) \right] 
\Phi\left({\textstyle\frac{1}{2}}\zeta\right) \right).
\label{Beq8}
\end{eqnarray}
Using the relation
\begin{equation}
\delta(y_0-{\textstyle\frac{1}{2}}\zeta_0)\left[ {T}_{0 \nu}(y), 
\Phi\left({\textstyle\frac{1}{2}}\zeta\right)
\right]=-i\delta^{(d)}({y}-{\textstyle\frac{1}{2}}\zeta)\frac{\partial \Phi(y)} 
{\partial y^{\nu}} 
\end{equation}
and substituting it into the Eq. (\ref{Beq8}) gives
\begin{eqnarray}
\frac{\partial}{\partial y_{\mu}}&& T \left( \Phi
{T}_{\mu \nu}
\Phi^{\dagger} \right)\nonumber\\
=&&-i\delta^{(d)}(y-{\textstyle\frac{1}{2}}\zeta) 
\left( \theta  (\zeta_0)
\frac{\partial \Phi(y)}  {\partial y^{\nu}}
\Phi^{\dagger}\left(-{\textstyle\frac{1}{2}}\zeta\right)+
\theta(-\zeta_0) \Phi^{\dagger}\left(-{\textstyle\frac{1}{2}}\zeta\right)
\frac{\partial \Phi(y)} 
{\partial y^{\nu}}\right) \nonumber \\
&&-i\delta^{(d)}(y+{\textstyle\frac{1}{2}}\zeta)
\left( \theta(\zeta_0)
\Phi\left({\textstyle\frac{1}{2}}\zeta\right)\frac{\partial \Phi^{\dagger}(y)} 
{\partial y^{\nu}} +
\theta(-\zeta_0) \frac{\partial \Phi^{\dagger}(y)} 
{\partial y^{\nu}} \Phi\left({\textstyle\frac{1}{2}}\zeta\right)
\right)\nonumber\\ 
=&&-i\delta^{(d)}(y-{\textstyle\frac{1}{2}}\zeta) 
\frac{\partial}{\partial y^\nu}\,
T\left(\Phi(y) \Phi^{\dagger}
\left(-{\textstyle\frac{1}{2}}\zeta\right)\right)-i 
\delta^{(d)}(y+{\textstyle\frac{1}{2}}\zeta)
\frac{\partial}{\partial y^\nu}\,T\left(
\Phi\left({\textstyle\frac{1}{2}}\zeta\right) \Phi^{\dagger}(y)\right)\, .
\label{Beq8a}
\end{eqnarray}
Substituting this expression into (\ref{Beq7}), integrating over $y$, and
then integrating by parts over $\zeta$ gives 
\begin{eqnarray}
q^{\mu} \Delta_M(p^{\prime}) {\cal J}_{\mu \nu}(p^{\prime}, p)\Delta_M(p)=&&
i\int d^d \zeta   e^{-ip \cdot \zeta}\; p^\nu <0| T\left( \Phi
\left({\textstyle\frac{1}{2}}\zeta\right)
\Phi^{\dagger}\left(-{\textstyle\frac{1}{2}}\zeta\right)
\right)|0> \nonumber\\
 &&-i\int d^d \zeta  
e^{-ip' \cdot \zeta}\; p'^\nu <0| T\left( \Phi
\left({\textstyle\frac{1}{2}}\zeta\right) \Phi^{\dagger}
\left(-{\textstyle\frac{1}{2}}\zeta\right) \right)|0>
\, .
\label{Beq7aa}
\end{eqnarray}
Finally, using the definition of the propagator [with the sign convention
of Eq.~(\ref{1eq6})]
\begin{equation}
i \int d^d \zeta e^{i p^{\prime} \cdot \zeta} 
<0|T \left( \Phi\left({\textstyle\frac{1}{2}}\zeta\right) 
\Phi^{\dagger}\left(-{\textstyle\frac{1}{2}}\zeta\right)
\right)|0>=\Delta_M(p^{\prime})\, ,
\end{equation}
Eq.~(\ref{Beq7aa}) becomes:
\begin{equation}
q^{\mu} \Delta_M(p^{\prime})  
{\cal J}_{\mu \nu}(p', p) \Delta_M(p)=
p_{\nu} 
\Delta_M (p)-p'_{\nu} \Delta_M(p') .
\label{Beq9}
\end{equation}
Canceling $\Delta(p^{\prime})$ 
and $\Delta_M(p)$ from the LHS of Eq.~(\ref{Beq9}), 
we get the gravitational Ward-Takahashi
identity for any bosonic field:
\begin{equation}
q^{\mu}{\cal J}_{\mu \nu}(p', p)=
p_{\nu} \Delta_M^{-1}(p')-p'_{\nu} 
\Delta_M^{-1}(p) \, .
\label{Beq10}
\end{equation}
Our proof of the Eq. 
(\ref{Beq10}) is dimension independent.

\begin{figure}
\centerline{\epsfxsize=1.5in\epsffile{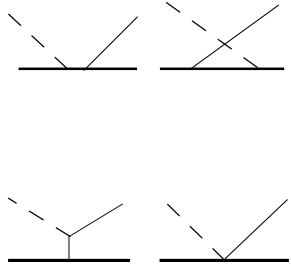}}
\vskip 0.2in
\caption{The $\Phi+\phi\to \Phi+h$ interaction to lowest order in
$\lambda$.}\label{four}
\end{figure}

\begin{figure}
\centerline{\epsfxsize=1.7in\epsffile{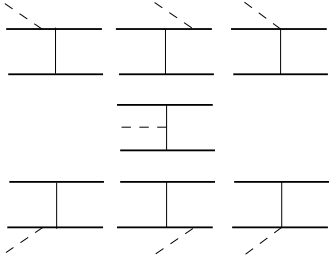}}
\vskip 0.2in
\caption{The radiative OBE diagrams to second order in $\lambda$.}\label{five}
\end{figure}

Similarly we can repeat this 
proof for any fermionic field, 
but we would have some sign 
differences in Eq. (\ref{Beq8}) 
and we would need to use anticommutation relations to 
redcuce the commutators to delta function. 
Hence the proof becomes model independent.
Next we will investigate the gauge 
invariance of some other diagrams.

\subsection{Gauge Invariance of 
Some Three-, Four- and Five-Point Functions}

Using the Ward-Takahashi identities we have derived, it is easy to show that 
the contributions from the four diagrams shown in Fig.~\ref{four} describing the
scattering of $\Phi+\phi\to \Phi+h$ (where $h$ is the gravitation) are gauge
invariant.

Similarily, the seven OBE (one boson 
exchange) diagrams shown in Fig.~\ref{five}, that describe graviton emmession
from the lowest order $\phi$ exchange contribution to $\Phi\Phi$ scattering 
are also gauge invariant.

\begin{figure}
\centerline{\epsfxsize=3.0in\epsffile{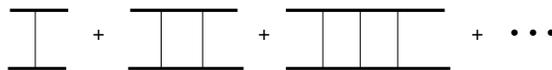}}
\vskip 0.2in
\caption{The $\phi$ ladder diagrams used to construct a $\Phi\Phi$
bound state.}\label{six}
\end{figure}

\begin{figure}
\centerline{\epsfysize=0.5in\epsffile{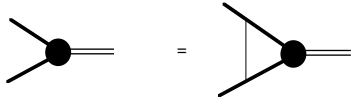}}
\vskip 0.2in
\caption{The ladder Bethe-Salpeter equation for the $\Phi\Phi$ bound
state.}\label{eight}
\end{figure}

\begin{figure}
\centerline{\epsfxsize=4in\epsffile{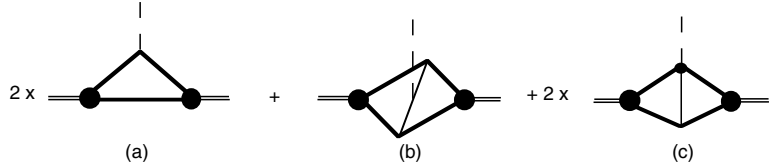}}
\vskip 0.2in
\caption{The gravitional coupling to the $\Phi\Phi$ bound state
in Bethe-Salpeter ladder approximation.}\label{nine}
\end{figure}

To investigate the gauge invariance of the gravitational coupling to the 
bound state of two $\Phi$-particles, we must first
define the Bethe-Salpeter vertex for the $\Phi\Phi$ bound state. In the
ladder approximation, the bound  state propagator is generated from the infinite
ladder sum (illustrated in Fig.~\ref{six}), and the bound state  vertex is
a solution of the ladder Bethe-Salpeter  equation, illustrated in
Fig.\ref{eight}.  With this description, the gravitional coupling to the
$\Phi\Phi$ bound state is described exactly by the 
three diagrams shown in Fig.~\ref{nine}, and is gauge invariant.  The diagram 
8(a) containing no internal 
$\phi$ lines and the diagram 8(c) 
which involves a four-particle vertex 
contribute twice (once for each $\Phi$ particle) and their sum is 
gauge invariant by itself. The diagram 8(b) 
contributes only once and is also separately gauge 
invariant. The proof follows from the Ward-Takahashi identies and 
does not depend on the explicit form of the propagators or bound
state vertices.  In a similar fashion
any $\phi^3$ theory, such as QED, is gauge invariant\cite{RG}.

How can we demonstrate the gauge invariance of a generic $\varphi^4$-type
theory?  No separate proof is necessary 
since the $\varphi^3$-type theories are 
gauge invariant and from any $\Phi^{\dagger} 
\Phi \phi$ model a generic 
$\varphi^4$-type theory can be obtained 
directly  by making the $\phi$ fields heavy and 
shrinking the internal $\phi$ propagators 
to a point (in this example the theory will have a point-like $(\Phi^{\dagger} 
\Phi)^2$ structure). In this way we
obtain  a effective field theory without loss of gauge invariance. 
After this reasoning we can assume that all 
$\varphi^4$-type theories are gravitionally gauge invariant.
   
These examples illustrate gravitational gauge invariance explicitly.
All of the diagrams of a given order, and certain infinite classes of
diagrams,  are gravitionally gauge invariant.
A consequence of this is that the gravitational field must couple to the total
energy-momentum  tensor of a composite system, and this coupling, which depends
on the bound state mass, must be consistent with other ways of computing the
bound state mass. In the following sections we will study the implications of
this consistency requirement for an effective $\varphi^4$-type field theory of
gravitation.

\subsection{Gravitational Coupling in an Effective Field Theory}

In this subsection we work out details of the gravitational coupling for
an effective field theory (EFT) with a $(\Phi^{\dagger} \Phi)^2$ structure.  
As discussed above, this EFT is obtained from the $\Phi^{\dagger} \Phi\phi$
theory by letting the mass $m$ of the $\phi$ particle go to
infinity, and shrinking all $\phi$ propagators to a point. This EFT would be
a valid approximation if the $\Phi$ particles have low momenta  and the $\phi$
particles are very heavy. 
 
\begin{figure}
\centerline{\epsfxsize=0.51in\epsffile{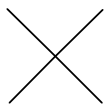}}
\vskip 0.2in
\caption{The $\Phi$ four-point coupling.}\label{ten}
\end{figure}

\begin{figure}
\centerline{\epsfxsize=0.51in\epsffile{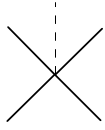}}
\vskip 0.2in
\caption{The $\Phi$-graviton five-point coupling.}\label{eleven}
\end{figure}

\begin{figure}
\centerline{\epsfxsize=2in\epsffile{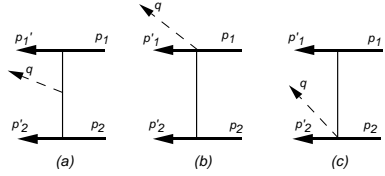}}
\vskip 0.2in
\caption{The three OBE $\Phi$-graviton 
diagrams which give the effective $\Phi$-graviton five-point
coupling illustrated in Fig.~10.}\label{twelve}
\end{figure}

This theory involves three couplings: an effective $(\Phi^{\dagger} \Phi)^2$
four point coupling, a $\Phi^{\dagger}\Phi h$ coupling, and a new effective 
$(\Phi^{\dagger} \Phi)^2\,h$ interaction.  The first two vertices were
discussed previously.  The effective $(\Phi^{\dagger} \Phi)^2$ coupling,
derived from the OBE model discussed above by shrinking the $\phi$ propagator
to a point, is 
\begin{equation}
-i{\cal V}_4=i\frac{\lambda^2}{m^2}=ig\, ,
\label{1eq14}
\end{equation}
where $g$ is the effective $\varphi^4$ coupling illustrated 
in Fig.~\ref{ten}. According to the relation (\ref{1eq11a}), 
this vertex implies the existence of 
a $(\Phi^{\dagger} \Phi)^2\,h$ five-point coupling of the form:
\begin{equation}
-i {\cal V}^e_{\mu \nu}=i\eta_{\mu\nu}{\cal V}_4=-i \eta_{\mu \nu}
\frac{\lambda^2} {m^2}=-i \eta_{\mu\nu} g \, .
\label{1eq15}
\end{equation}
which is shown in Fig.~\ref{eleven}.
On the other hand, consistency 
requires that this five-point vertex also be 
obtained by shrinking the three 
diagrams shown in Fig.~\ref{twelve}.

To prove this consistency, we compute each of the 
diagrams in Fig.~\ref{twelve}.  Diagram \ref{twelve}(a), in the limit of low
external  momenta and low momentum 
transfer, becomes:
\begin{eqnarray}
-i{\cal M}_a
=&&\lambda^2 \Delta_m(k) (-i 
{\cal J}^m_{\mu \nu}) \Delta_m(k-q)  \\ \nonumber
\simeq&& \frac{\lambda^2}{m^2} (i \eta_{\mu \nu} 
m^2) \frac{1}{m^2} \\ \nonumber
\simeq&& i \eta_{\mu \nu} \frac{\lambda^2}{m^2} \, ,
\label{1eq15a}
\end{eqnarray}
where $k=p_1-p_1^{\prime}$ and $q$ is the 
momentum of the outgoing graviton. Note that, at small momenta,
Eq.~(\ref{1eq10}) gives $-{\cal J}^m_{\mu \nu} \simeq\eta_{\mu \nu} m^2$.  The 
contribution from the second diagram, \ref{twelve}(b), is:
\begin{eqnarray}
-i {\cal M}_b=&& -i \Delta_m(k-q) (i \lambda \eta_{\mu \nu}) 
(-i \lambda)   \\ \nonumber
\simeq&& -i\eta_{\mu \nu} \frac{\lambda^2}{m^2}\,  .    
\label{1eq15b}
\end{eqnarray} 
The contribution from diagram \ref{twelve}(c) in the low momemtum limit is
equal to \ref{twelve}(b), and hence the total contribution from diagrams
(a)--(c) is 
\begin{equation}
-i ({\cal M}_a+{\cal M}_b+{\cal M}_c) 
\simeq  -i  \eta_{\mu \nu} \frac{\lambda^2}{m^2}=-i \eta_{\mu\nu} g\, ,
\label{1eq15d}
\end{equation}
in full agreement with the Eq.~(\ref{1eq15}).   We have shown that the
five-point vertex of EFT emerges from the low momentum limit of the ``full''
theory.  We will now use this understanding to obtain an effective description
of the bound state.

In the following discussion, we will skip back and forth between the 
``exact'' theory (which includes the OBE description of the interaction) and the
EFT (in which the interaction is contracted to a point).  We emphasize that
these theories are equivalent at small momentum. 

\begin{figure}
\centerline{\epsfxsize=2.77in\epsffile{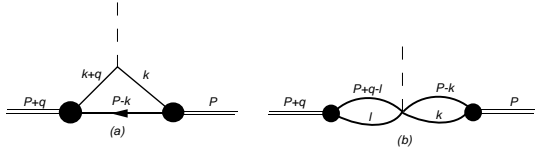}}
\vskip 0.2in
\caption{Graviton coupling to the bound state in the $(\Phi^{\dagger}\Phi)^2$
EFT.} \label{twenty}
\end{figure}

At low momenta, the dressed $\Phi\Phi$ scattering amplitude 
can be calculated in two ways that are equivalent at low momenta: by adding the
ladder diagrams of Fig.~6, or by summing the bubbles resulting from the
effective four-point interaction illustrated in Fig.~\ref{ten}.  In either
case, the infinite sum generates the bound state.  The gravitational
interaction with the bound state can also be described in two different ways,
which are equivalent at low momenta.  The first of these, illustrated in
Fig.~\ref{nine}, is the description in the ``exact'' theory.  The second,
shown in Fig.~\ref{twenty}, is the description in the EFT.

In either theory, the momentum conservation implied by the gauge invariance of
the gravitational interaction implies that the effective graviton-bound
state interaction at low momentum transfer must be proportional to the 
energy momentum tensor of the bound state
\begin{equation}
-i{\cal J}_{M_b}^{\mu \nu}(p',p)
=i\left[p^{\mu} p^
{\prime \nu}+p^{\nu} p^{\prime \mu}
+g^{\mu \nu}(M_b^2-p \cdot p^{\prime})\right]\, .
\label{1eq16}
\end{equation}
The interactions shown in Figs.~\ref{nine} and \ref{twenty} {\it must reproduce
this result as the momentum of the graviton $q\to0$\/}.  This requirement
places a constraint in the bound state wave function.  As we have shown, it
is sufficient to study this constraint in the context of the EFT, and
this is the subject of the following sections.

\section{The Mass of a Composite System of Scalars}

Using the EFT discussed in the previous section, the tensor
(\ref{1eq16}) must be the sum of the
two contributions shown in Fig.~\ref{twenty}
\begin{equation}
-i{\cal J}_{M_b}^{\mu \nu}(P+q,P) = -2i{\cal J}_a^{\mu \nu}(P+q,P) -i{\cal
J}_b^{\mu\nu}(P+q,P) \, , \label{totalj}
\end{equation}
where ${\cal J}_a^{\mu \nu}$ and ${\cal J}_b^{\mu \nu}$ are the contributions
from diagrams \ref{twenty}(a) and \ref{twenty}(b), respectively.  The factor
of 2 multiplying ${\cal J}_a$ displays the fact that it occurs twice, once
for each constituent.  These diagrams are 
\begin{equation}
-i{\cal J}_a^{\mu \nu}(P+q,P)
=-{\cal N}^2\int \frac{d^d k}{(2 \pi)^d} \Delta_M(P-k) 
\Delta_M(k+q) {\cal J}^{\mu \nu}_M(k+q,k) \Delta_M(k)\,  ,
\label{2eq1bb}
\end{equation}
where ${\cal N}$ as the bound state vertex function (constant in the
EFT), and 
\begin{equation}
-i{\cal J}_b^{\mu \nu}(P+q,P)
=ig \eta^{\mu \nu} {\cal N}^2 b\left((P+q)^2\right) b(P^2) \, ,
\label{2eq1c}
\end{equation}
where, in 1+1 dimension, the bubble $b$ is
\begin{equation}
b(P^2)=i\int \frac{d^d k}{(2 \pi)^d} 
\frac{1} {(M^2-k^2)[M^2-(P-k)^2]}\, ,
\label{2eq1d}
\end{equation}
as given in Ref.~\cite{ZFBG1}.

In Ref.~\cite{ZFBG1} we showed that the vertex function (or normalization 
constant) ${\cal N}$ and the coupling constant $g$ are related to the
bound state mass by
\begin{eqnarray}
{\cal N}=\left(-\frac{\partial b(M_b^2)}
{\partial M_b^2} \right)^{-1/2} \qquad\qquad g=\frac{1}{b(M_b^2)}\, ,
\label{2eq4}
\end{eqnarray}
Requiring the initial and final bound states to be on shell, so the
$P^2=M_b^2$ and $(P+q)^2=M_b^2$, permits us to rewrite Eq.~(\ref{2eq1c}):
\begin{equation}
-i{\cal J}_b^{\mu \nu}=i \eta^{\mu \nu}{\cal N}^2 b(M_b^2) \, .
\label{2eq1d1}
\end{equation}

Before we use Eq.~(\ref{totalj}) to calculate the bound state mass, we will
prove that this expression is gauge invariant. First, compute $-iq_{\mu}{\cal
J}_a^{\mu\nu}$
\begin{eqnarray}
-iq_{\mu}{\cal J}_a^{\mu \nu}&&(P+q,P)\nonumber\\
=&&-{\cal N}^2\int \frac{d^d k}{(2 \pi)^d} \Delta_M(P-k) 
\Delta_M(k+q) q_{\mu}
{\cal J}^{\mu \nu}_M(k+q,k) \Delta_M(k) \nonumber\\
=&&-{\cal N}^2 \int \frac{d^d k}{(2 \pi)^d} k^{\nu} 
\Delta_M(P-k) \Delta_M(k) 
+{\cal N}^2 \int \frac{d^d k}{(2 \pi)^d} 
(k+q)^{\nu} \Delta_M(P-k) 
\Delta_M(k+q)  \nonumber \\
=&&-i q^{\nu} {\cal N}^2 \frac{1}{2}\, b(M_b^2) \, . 
\label{2eq1e}
\end{eqnarray}
The transition from the second line was done by using the Ward-Takahashi
identity.  In the last step we symmetrized the integrands ($k\to k+P/2$
in the ifrst integral and $k\to k+(P-q)/2$ in the second), droped the odd
$k^\nu$ term, and extracted the final result using the definition
(\ref{2eq1d}). Recall that this term must be multiplied by 2, and is
cancelled exactly by the term from Eq.~(\ref{2eq1d1}), which is
\begin{equation}
-iq_{\mu}{\cal J}_b^{\mu \nu}=iq^{\nu}{\cal N}^2 b(M_b^2) \, .
\label{2eq1d2}
\end{equation}
Hence, the sum of the diagrams from Fig.~\ref{twenty} 
is gauge invariant. 

We now use Eq.~(\ref{totalj}) to ``calculate'' the square of the 
bound state mass in 1+1 dimension.   Setting $q=0$ and contracting the indices
in  the Eq. (\ref{1eq16}) projects out the bound state mass:
\begin{equation}
\eta_{\mu \nu} {\cal J}_{M_b}^{\mu \nu}(P,P)=-2 M_b^2\, .
\label{2eq1a}
\end{equation}
Our goal is to demonstrate that Eq.~(\ref{totalj}) is indeed consistent with
the above result.

First notice that, for off-shell particles in 1+1 $d$,
\begin{eqnarray}
\eta_{\mu\nu} {\cal J}_{M}^{\mu \nu}(k,k) 
\Delta_M^2(k)=&& 2 M^2
\frac{\partial \Delta_M(k)}{\partial M^2}  \, 
\label{2eq1}
\end{eqnarray}
and hence the contraction of Eq.~(\ref{totalj}) gives
\begin{eqnarray}
-i \eta_{\mu \nu}{\cal J}_{M_b}^{\mu \nu}(P,P) =&& 2iM_b^2 \nonumber\\
=&&-2{\cal N}^2\int \frac{d^d k}{(2 \pi)^d} \Delta_M(P-k) 
 2M^2 \frac{\partial \Delta_M(k)}{\partial M^2} 
+2i {\cal N}^2 b(M_b^2) \\ \nonumber
=&&2i{\cal N}^2 \left( M^2 \frac{\partial}{\partial M^2}+1 \right) b(M_b^2)\, .
\label{2eq2}
\end{eqnarray}
Replacing ${\cal N}^2$ by the normalization condition (\ref{2eq4}) allows us
to transform this result into the following condition
\begin{equation}
\left[M^2_b \frac{\partial}
{\partial M_b^2}+M^2\frac{\partial}
{\partial M^2}+1 \right] b(M_b^2)=0\, .
\label{2eq5}
\end{equation}
To verify the validity of Eq. (\ref{2eq5}), we recall that the explicit form
of $b$ (taken from Ref.~\cite{ZFBG1}, with $m_1=m_2=M$) is
\begin{equation}
b(M_b^2)=-\frac{1}{4 \pi} \int \frac{dx}{M^2-M^2_bx(1-x)}\, . 
\label{2eq4a}
\end{equation}
This satisfies the condition (\ref{2eq5}).  
Consequently, we have shown that the {\it normalization condition\/},
Eq.~(\ref{2eq1c}), {\it also insures that the graviton couples to the total
mass of the bound state, as required by energy- momentum conservation\/}.  Since
the normalization condition originally came from baryon number conservation (as
discussed in Ref.~\cite{ZFBG1}), we conclude that {\it energy-momentum
conservation and baryon number conservation are different aspects of the same
constraint\/}.

\section{Bound State of a Spin $1/2$ Fermion and a Scalar Boson}

In this section we extend the preceding results to the case of a bound state
of a scalar and a spin $1/2$ fermion, still in $1+1$ dimensions.  As above,
we assume an effective four-point contact interaction of the form
$\lambda(\Psi^\dagger\Psi)(\Phi^\dagger\Phi)$, as described in
Ref.~\cite{ZFBG1}.   We start by reviewing the results of the bound state
calculations from Ref.~\cite{ZFBG1}, and then describe the graviton-spinor
interaction.  We conclude the section with a detailed treatment of the bound
state.

\subsection{Bound State Formalism and Normalization}

The scattering amplitude (droping the initial and final spinors) for this
$\varphi^4$-type interaction depends only on the total momentum, $p$, and can
be written
\begin{eqnarray}
{\cal M}=&&\frac{\lambda}{1-\lambda\left(A(p^2)+\not{p}B(p^2)\right)}
\nonumber\\
=&&\frac{\lambda (1- \lambda A(p^2) + \lambda B(p^2) \not{p})}{
(1-\lambda A(p^2))^2-\lambda^2B^2(p^2)p^2}\, ,
\label{3eq1} 
\end{eqnarray}
where the spin 1/2 bubble is $A(p^2)+B(p^2) \not{p}$, and the functions
$A(p^2)$ and $B(p^2)$ are
\begin{eqnarray}
A(p^2)=-\frac{m_1}{4 \pi} \int^1_0 \frac{dx}
{m^2_1x+m^2_2(1-x)-p^2x(1-x)} \nonumber \\
B(p^2)=-\frac{1}{4 \pi} \int^1_0 
\frac{(1-x)dx}{m^2_1x+m^2_2(1-x)-p^2x(1-x)}\, ,
\label{3eq2}
\end{eqnarray}
with $m_1$ the mass of the fermion and $m_2$ the mass of the boson.
If the bound state mass is $M_b$, the scattering amplitude 
${\cal M}$ should have a pole at $p^2=M_b^2$, which implies
\begin{equation}
1-\lambda A(M_b^2)=\lambda M_b B(M_b^2)\, .
\label{3eq3}
\end{equation}
Using this condition to fix $A(M_b^2)$, expanding 
both the numerator and denominator of (\ref{3eq1})
in powers of $p^2-M_b^2$,
and keeping the lowest order terms only, gives
\begin{equation}
{\cal M} \simeq -{\cal N}^2 \frac{M_b+ \not{p}}
{M_b^2-p^2}=-\frac{{\cal N}^2}{M_b- \not{p}}\, ,
\label{3eq4}
\end{equation}
where
\begin{equation}
{\cal N}=\frac{1}{\sqrt{-2M_b(A^{\prime}
(M_b^2)+M_b B^{\prime}(M_b^2))-B(M_b^2)}} \, ,
\label{3eq5}
\end{equation}
is the bound state normalization constant or the bound state vertex function
(which is again a constant in this model). The prime in Eq.~(\ref{3eq5}) denotes
the derivative with respect to $M_b^2$.
Using the condition (\ref{3eq5}), the equations (\ref{3eq2})
for $A$ and $B$, and the identity
\begin{equation}
\int^1_0 dx \frac{m_1^2x^2-m_2^2(1-x)^2}
{[m_1^2x+m_2^2(1-x)-M_b^2x(1-x)]^2}=0
\label{3eq7}
\end{equation}
we can rewrite the normalization 
condition (\ref{3eq5}) as:
\begin{equation}
\frac{{\cal N}^2}{4 \pi} 
\int^1_0 dx \frac{x \left[ M_b(1-
x)+m_1 \right] ^2}
{[m_1^2x+m_2^2(1-x)-M_b^2x(1-x)]^2}=1.
\label{3eq8}
\end{equation}
We use this form to show that energy-momentum conservation 
is compatible with the normalization condition.

\subsection{Interaction of Gravitons and Spinors}

A rigorous description
of the interaction of the graviton with 
spinors can be found in Ref.~\cite{DES}. This description 
uses the  concept of a vierbein or tetrad, ${\bf e}_{\mu}$, related to the
metric tensor as follows:
\begin{equation}
{\bf e}_{\mu}(x) \cdot {\bf e}_{\nu}(x)=
e^a_{\mu}(x) e^a_{\nu}(x)=g_{\mu \nu}(x)\, .
\end{equation} 
Likewise the Dirac matrices in curved space $\gamma^{\mu}(x)$
are related 
to the Dirac matrices in
flat space ${\tilde \gamma}^a$ via: 
\begin{equation}
\gamma^{\mu}(x)=e^{a \mu}(x) {\tilde \gamma}^a\, ,
\end{equation}
where $e^{a \mu}(x)$ and $e^a_{\mu}(x)$ are the inverse of each other:
\begin{equation}
e^{a \mu}(x) e^b_{\mu}(x)=\delta^{a b}\, .
\end{equation} 
In the presence of gravity, the flat space Dirac matrices in the Lagrangian
are  replaced by the $\gamma^{\mu}(x)$, and  normal derivatives are 
replaced by covariant derivatives (involving the affine connections).   
The covariant derivatives are no longer equal to the normal derivatives, as in
the scalar case. Finally, the action is obtained by integrating the resulting
Lagrangian density, multiplied by $\sqrt{-g}=\det[e^a_{\mu}]$, over
$x$.  From this action one is able to derive the 
Feynman rules by varying the symmetric part of the perturbation of 
the vierbein, which is the gravitational field.

However, since such a rigorous treatment would be very long and difficult, 
here we will give a less rigorous and simpler,  more intuitive
development.  Our discussion starts with the observation that the graviton
should couple to the  energy-momentum tensor of the Dirac field.  This tensor
is easily calculated, and would lead to the following Feynman rule for the
graviton coupling to a spin 1/2 particle:
\begin{equation}
-i{\cal J}_0^{\mu \nu}(p',p)=\frac{i}{2} \left[ 
\gamma^{\mu}(p+p^{\prime})^{\nu}-\eta^{\mu \nu}
(\not{p}^{\prime}+ \not{p}-2 m_1) \right]\, .
\label{3eq9}
\end{equation}
Here $p$ is the momentum of the incoming 
fermion, $p^{\prime}$ is the momentum of the outgoing fermion,
and $q=p^{\prime}-p$ is the wave 
number of the graviton. 
This Feynman rule is symmetric in 
the in- and outgoing fermion 
momenta, but not in the Lorentz 
indices, so it must be fundamentally wrong.  Its only virtue is  
that it satisfies the Ward-Takahashi identity in the first index $\mu$:
\begin{equation}
q_{\mu}{\cal J}_0^{\mu \nu}(p',p)=
\left[ p^{\nu} S_{m_1}^{-1}(p') - 
p'^{\nu} S_{m_1}^{-1}(p) \right]\, ,
\label{3eq10}
\end{equation}
where the fermion propagator is
\begin{equation}
S^{-1}_{m_1}(p)=m_1- \not{p}\, .
\label{3eq11}
\end{equation}
To get around this problem, 
we correct our former Feynman rule by adding to it another 
term $t^{\mu \nu}(p',p)$:
\begin{equation}
{\cal J}_0^{\mu \nu}(p',p) 
\rightarrow {\cal J}^{\mu \nu}(p',p)=
{\cal J}_0^{\mu \nu}(p',p)+t^{\mu \nu}(p',p)\, .
\label{3eq11a}
\end{equation}
The following conditions will uniquely fix the
additional term $t^{\mu\nu}(p',p)$:
\begin{itemize}
\item ${\cal J}^{\mu \nu}$ must be symmetric in its Lorentz indices, 

\item  ${\cal J}^{\mu \nu}$ must satisfy the Ward-Takahashi identity in
both indices (implying that $t^{\mu\nu} (p',p)$ must be divergenceless in the
first index), 

\item ${\cal J}^{\mu \nu}$ must be Hermitian, implying that $\gamma^0
[{\cal J}^{\mu\nu}(p',p)]^T \gamma^0={\cal J}^{\nu \mu}(p,p')$, 

\item $t^{\mu \nu}(p',p)$ must be traceless so that the mass, which is
projected out by taking the trace of ${\cal J}^{\mu \nu}$, is unmodified. 
\end{itemize}

\noindent These conditions lead to the following {\it unique\/} form for 
$t^{\mu\nu}$:    
\begin{eqnarray}
t^{\mu \nu}(p',p)
=&&-\frac{1}{2} 
\left[ \gamma^{\nu}\left( P^{\mu}-
\frac{P \cdot q}{q^2} q^{\mu} \right)
-\frac{P \cdot q}{q^2} q^{\nu} 
\gamma^{\mu} +
P \cdot q \not{q} \frac{q^{\mu} q^{\nu}}{q^4} \right] \\ \nonumber  
&&+\frac{1}{4} 
\left(P^{\mu}-\frac{P \cdot q}{q^2} 
q^{\mu} \right) 
\left(\gamma^{\nu}-\frac{\not{q}}{q^2} 
q^{\nu} \right) \\ \nonumber 
&&+ \frac{1}{4} \left( P^{\nu}-
\frac{P \cdot q}{q^2} q^{\nu} \right)
\left( \gamma^{\mu}-\frac{\not{q}}{q^2} q^{\mu} \right) ,
\label{3eq11b}
\end{eqnarray}
with $P=p+p^{\prime}$. 

Because the Ward-Takahashi identity 
now holds for both indices, our formal 
investigation of scalar particles can be 
repeated step by step.  In particular, since the proof of gauge invariance
relied on the Ward-Takahashi identities, it should hold here as well.

Since the bound state of a spin $1/2$ fermion and a scalar boson  
is also a spin $1/2$ fermion, we expect 
the graviton-bound state interaction to be also given by
(\ref{3eq11a}). Replacing $m_1$ by $M_b$, and letting $q\to0$, gives the
following relation for the bound state mass in 1+1 dimensions
\begin{equation}
\eta_{\mu \nu}{\cal J}^{\mu \nu}(p,p)=
-\left[M_b + S_{M_b}^{-1}(p)\right]\to -M_b\, , 
\label{3eq11c}
\end{equation}
where the last relation follows for on-shell particles, as discussed below.

\subsection{Mass of a Composite Fermion}

\begin{figure}
\centerline{\epsfxsize=2.46in\epsffile{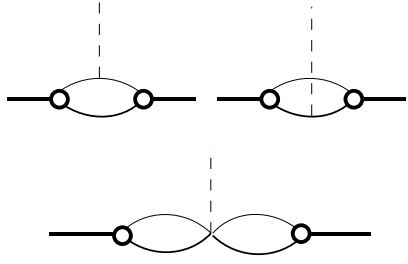}}
\caption{Interaction of the graviton with a bound state of two unequal mass
particles in the EFT. }\label{twentyone}
\end{figure}

As in the scalar EFT, the three diagrams shown in Fig.~\ref{twentyone} give
the EFT interaction of a graviton with the spin 1/2 bound state described
above.  [Since the two constituent particles have different
properties, the single diagram \ref{twenty}(a) now becomes the first two
diagrams shown in Fig.~\ref{twentyone}.]   In the limit when $q \rightarrow
0$, the bound state mass can be obtained from
\begin{eqnarray}
\eta_{\mu \nu}{\cal J}_{M_b}^{\mu \nu}
(p,p)=&&-{\bar {\cal U}}(p)
\left[M_b+S^{-1}_{M_b}(p)\right]{\cal U}(p)= \nonumber \\
=&&-M_b \,{\bar {\cal U}}(p){\cal U}(p)=-M_b\, .
\label{3eq13}
\end{eqnarray}

Using the relation (\ref{2eq1}) and the similar 
relation
\begin{eqnarray}
S_{m_1}(k)\,\eta_{\mu \nu}{\cal J}_{m_1}^{\mu \nu}(k,k)\,S_{m_1}(k)
=-m_1\,S^2_{m_1}(k)-S_{m_1}(k)
=\left(m_1\frac{d}{dm_1} -1\right)S_{m_1}(k)
\, , \label{3eq14}
\end{eqnarray}
the first two Feynman diagrams in Fig.~\ref{twentyone} become,
as $q \rightarrow 0$,
\begin{eqnarray}
\eta_{\mu \nu}{\cal J}_{1+2}^{\mu \nu}
(p,p)=&& -i{\bar {\cal U}}(p){\cal N}^2\int \frac{d^d k}{(2 \pi)^d} \Bigl\{
\Delta_{m_2}(p-k)\left(m_1\frac{d}{dm_1} -1\right)S_{m_1}(k)\nonumber\\
 &&\qquad + S_{m_1}(p-k)\, 2
m_2^2\frac{d}{dm_2^2} \Delta_{m_2}(k)\Bigr\}{{\cal U}}(p) \nonumber\\
=&&-{\bar {\cal U}}(p){\cal N}^2 \left( -1+m_1 
\frac{\partial}{ \partial m_1}+2m_2^2 
\frac{\partial}{\partial m_2^2}
\right) \left( A(M_b^2)+\not{p}  
B(M_b^2) \right){\bar {\cal U}}(p)\nonumber\\
=&& -{\cal N}^2 \left( -1+m_1 
\frac{\partial}{ \partial m_1}+2m_2^2 
\frac{\partial}{\partial m_2^2}
\right) \left( A(M_b^2)+M_b  
B(M_b^2) \right) \, .
\label{3eq15}
\end{eqnarray}
As in the scalar case, the third diagram becomes
\begin{eqnarray}
\eta_{\mu \nu}{\cal J}_{3}^{\mu \nu}
(p,p)=&&-2\lambda {\cal N}^2 {\bar {\cal U}}(p)\left(A(M_b^2)+\not{p}  
B(M_b^2) \right)^2 {{\cal U}}(p) \nonumber\\
=&&-2\lambda {\cal N}^2 \left(A(M_b^2)+M_b 
B(M_b^2) \right)^2 = -2 {\cal N}^2 \left(A(M_b^2)+M_b 
B(M_b^2) \right)\, , \label{3eq15aa} 
\end{eqnarray}
where Eq.~(\ref{3eq3}) was used in the last step.  Adding (\ref{3eq15}) and
(\ref{3eq15aa}) and equating them to (\ref{3eq13}) gives the following
relation  
\begin{equation}
M_b ={\cal N}^2 \left( 1+m_1 
\frac{\partial}{ \partial m_1}+2m_2^2 
\frac{\partial}{\partial m_2^2}
\right) \left( A(M_b^2)+M_bB(M_b^2) \right)\, .
\label{3eq16}
\end{equation}
Substituting the expressions (\ref{3eq2}) for $A$ 
and $B$ into this equation, and using the identity (\ref{3eq7}), 
transforms the equation into
\begin{equation}
M_b=\frac{{\cal N}^2}{4 \pi} 
\int^1_0 \frac{M_b\, x \left[ M_b(1-x)+m_1
\right] ^2dx}{\left[m_1^2x+m_2^2(1-x)-
M_b^2x(1-x) \right]^2}\,
\label{3eq17}
\end{equation}
which is Eq.~(\ref{3eq8}).  Consequently the bound state normalization 
condititon insures the conservation of energy-momentum in this example as well.

\section{CONCLUSIONs}

We have shown for two specific examples in 1+1 dimension that the bound state
normalization condition and the energy-momentum conservation condition
are, in fact, identical constraints.  Similarily, it has been
shown\cite{ZFBG1} for the same systems (and it is true in general) that the
normalization condition and charge conservation (or baryon conservation) are
also equivalent.  For systems with no conserved vector current (charge or
baryon number) the conservation of the energy-momentum tensor current
associated with gravity may be of special significance.   

While our discussion did not apply to higher dimensions, where form factors or
cutoffs are needed to regularize the EFT, we {\it conjecture\/} that
energy-momentum conservation and the bound state normalization condition
must also be equivalent in these cases.

We conclude this paper by returning to our opening discussion of momentum
conservation in DIS.  While the interaction of a graviton with a bound state
would seem to be unrelated to DIS, our proof of energy-momentum conservation
ought to imply that the covariant discription of DIS conserves momentum.  In
fact, for the two body bound states in 1+1 dimension that we have been
discussing, the demonstration is very simple. 

In the scalar model investigated in Ref.~\cite{ZFBG1}, the bound state
consisted of a charged particle of mass $m_1$ and a neutral particle of mass
$m_2$.  The charge distribution function is 
\begin{equation}
f(x)=\frac{{\cal N}^2}{4 \pi} \frac{x(1-x)}{\left[m_1^2x +
m_2^2(1-x) -  M^2\, x(1-x)\right]^2} \, , 
\label{1eqlast}
\end{equation}
and charge conservation (and the bound state normalization
condition) lead to the requirement that
\begin{equation}
\int_0^1 dx  f(x)=1\, .
\label{0eq1a}
\end{equation}
Note that the distribution amplitude (\ref{1eqlast}) is symmetric under the
interchange of $x\to1=x$ and $m_1\to m_2$, so that if both $m_1$ and $m_2$
were charged, we could define two ``flavor'' distributions
\begin{eqnarray}
f_1(x)=&&f(x)\nonumber\\
f_2(x)=&&f(1-x)
\end{eqnarray}
both normalized to unity.  This would insure charge or baryon number
conservation
\begin{equation}
b_1\int_0^1 dx  f_1(x)+b_2\int_0^1 dx  f_2(x)=b_1+b_2
\label{0eq1a1}
\end{equation}
as in Eq.~(\ref{eq1aa}).  Note that these conditions also insure that momentum
is conserved.  In this notation, Eq.~(\ref{eq1}) is written
\begin{equation}
\int_0^1 dx\, x f_1(x)+\int_0^1 dx\, x f_2(x)=\int_0^1 dx\, x f_1(x)+\int_0^1 dx
\,(1-x) f_1(x)=1
\label{0eq1a2}
\end{equation}
in complete agreement with Eq.~(\ref{eq1}).  We conclude that momentum is
conserved in covariant models of DIS.

\section{acknowledgments}
This research has been supported in part by the DOE under grant
No.~DE-FG02-97ER41032, and by DOE contract DE-AC05-84ER40150 administered by
SURA in support of the Thomas Jefferson National Accelerator Facility.  One of
us (Z.B.) is a  recipient of the  graduate fellowship awarded by SURA and the
Thomas Jefferson  National Accelerator Facility.  Support provided by the DOE
and the SURA fellowship are gratefully acknowledged.  One of us (F.G.) wishes
to acknowledge interesting conversations on this topic with Geoffrey West at
the Institute for Nuclear Theory, Seattle Washington.

\end{document}